\documentstyle[prl,aps,preprint]{revtex}
\input psfig

\begin{document}

\widetext
\draft
%\twocolumn[\hsize\textwidth\columnwidth\hsize\csname @twocolumnfalse\endcsname
\title{
Renormalization of Fluctuating Tilted-Hexatic Membranes
}
\author{
Jeong-Man Park
}
\address{
Department of Physics, 
Catholic University of Korea, Seoul, Korea
}
\maketitle

\begin{abstract}
We consider the tilted-hexatic Hamiltonian on the fluctuating membranes.
A renormalization-group analysis leads us to find three critical
regions; two correspond to the strong coupling regimes of the gradient
cross coupling where we find the (anti-)locked tilted-hexatic to
liquid phase transition, the other to the weak coupling regime
where we find four phases; the unlocked tilted-hexatic phase, the hexatic 
phase, the tilted phase, and liquid phase. 
The crinkled-to-crumpled transition of the fluctuating tilted-hexatic 
membranes is also described. 
\end{abstract}
\pacs{PACS numbers: 05.70.Jk, 68.10.-m, 87.22.Bt}
%]
\clearpage
%\narrowtext
\par
Lyotropic liquid-crystal systems show a variety of phases with different
types of in-plane two dimensional order.
Among the most interesting are the tilted-hexatic phases, which have
quasi-long-range order in two order parameters (the orientation of the local
bond and the direction of the local molecular tilt), but only short-range
translational order.
Recently, there has been considerable progress in understanding 
tilted-hexatic phases on the rigid layered liquid-crystals.
Selinger and Nelson \cite{Selinger-88} have presented a Landau theory
for transitions among tilted-hexatic phases.
They consider the tilt-bond interaction potential and find several
different hexatic phases differing from each other
in the relation between the local bond orientation and the local tilt 
direction depending on the interaction potential parameters.
However, they consider only phase transitions 
between low temperature phases (tilted-hexatic phases) 
on the rigid 2-dimensional plane, in which disclinations
in the bond orientational angle field $\theta_{6}({\bf u})$ and vortices 
in the tilt-angle field $\theta_{1}({\bf u})$ 
can be neglected.
\par 
In this Letter, we present a Landau theory, without the tilt-bond interaction
potential, for transitions from tilted-hexatic phase to disordered liquid
phase on the fluctuating membranes.
We consider the fluctuating membranes with the tilt and the hexatic 
in-plane orders described by the order parameters $\psi_{1}=e^{i\theta_{1}}$
and $\psi_{6}=e^{i6\theta_{6}}$, respectively. 
The tilt and the hexatic orders are coupled to each other via
a gradient cross coupling introduced by Nelson and Halperin \cite{Nelson-80}.
Depending on this gradient cross coupling parameter, we find three critical
regions in the phase space of the tilt stiffness $K_{1}$, the hexatic
stiffness $K_{6}$, and the gradient cross coupling $K_{16}$; two correspond
to the strong coupling and the other the weak coupling.
We also show that without the tilt-bond interaction potential there
exist a couple of different tilted-hexatic phases.
Finally, we discuss the crumpling transition of the fluctuating
membranes with the tilt and the hextic in-plane orders.
\par
We parametrize the membrane by its position vector as a function of
standard Cartesian coordinates ${\bf x}=(x,y)$;
\begin{equation}
{\bf R}({\bf x})=({\bf x},h({\bf x})),
\end{equation}
where $h({\bf x})$ measures the deviation from the flat surface.
This is called a Monge gauge.
Associated with ${\bf R}({\bf x})$ is a metric tensor 
$g_{\alpha\beta}({\bf x})=
\partial_{\alpha}{\bf R}({\bf x})\cdot\partial_{\beta}{\bf R}({\bf x})$ 
and a curvature tensor $K_{\alpha\beta}({\bf x})$ defined via 
$K_{\alpha\beta}({\bf x})= {\bf N}({\bf x})\cdot\partial_{\alpha}
\partial_{\beta}{\bf R}({\bf x})$, 
where ${\bf N}({\bf x})$ is the local unit normal to the surface.
From the curvature tensor $K_{\alpha\beta}$, the mean curvature, $H$, and the 
Gaussian curvature, $K$, are defined as follows:
\begin{equation}
H = \frac{1}{2} g^{\alpha\beta}K_{\beta\alpha}, \;\; 
K = \det g^{\alpha\lambda}K_{\lambda\beta},
\end{equation}
where $g^{\alpha\beta}$ is the inverse tensor of $g_{\alpha\beta}$ 
satisfying $g^{\alpha\lambda}g_{\lambda\beta}=\delta^{\alpha}_{\beta}$.
In the continuum elastic theory, the long-wavelength properties
of a fluctuating membrane are described by 
the Helfrich-Canham Hamiltonian \cite{Helfrich-73}
\begin{equation}
{\cal H}_{\rm HC} = \frac{1}{2} \int d^{2}x \sqrt{g} \left(
\kappa H^{2} + \bar{\kappa} K + \sigma \right),
\end{equation}
where $g=\det g_{\alpha\beta}$, $\kappa$ is the bending rigidity,
$\bar{\kappa}$ is the Gaussian rigidity, and $\sigma$ is the tension of
the membrane.
The first term is the mean curvature energy,
the second the Gaussian curvature energy, and the third 
the surface tension energy. We are mostly interested in free membranes
for which the topology is fixed and
the renormalized surface tension obtained by differentiating
the total free energy ${\cal F}$ with respect to the total surface area
${\cal A}$ $(\sigma_{R} = \partial {\cal F}/\partial {\cal A})$, is zero.
Therefore, we will ignore the Gaussian curvature energy due to the topological
invariance (the Gauss-Bonnet theorem) \cite{Spivak-79}
and the surface tension energy 
with the understanding
that it is really present if we want to keep track of how $\sigma_{R}$
actually becomes zero \cite{David-91}.
In the Monge gauge, the mean curvature energy for the geometric shape 
fluctuations becomes
\begin{equation}
{\cal H}_{\rm HC} = \frac{1}{2}\kappa \int d^{2}x \sqrt{1+(\nabla h)^{2}}
\left[ \nabla \cdot\left( \frac{\nabla h}{\sqrt{1+(\nabla h)^{2}}} \right)
\right]^{2}.
\end{equation}
\par
We consider the Hamiltonian for the tilt and the hexatic orders on a 
fluctuating membrane \cite{Thesis-94},
\begin{eqnarray}
{\cal H}_{\rm TH} & = & \frac{1}{2} K_{1} \int d^{2}x\sqrt{g}g^{\alpha\beta}
(\partial_{\alpha}\theta_{1}-A_{\alpha})(\partial_{\beta}\theta_{1}-A_{\beta}) 
\nonumber \\
& + & \frac{1}{2} (36K_{6}) \int d^{2}x\sqrt{g}g^{\alpha\beta}
(\partial_{\alpha}\theta_{6}-A_{\alpha})(\partial_{\beta}\theta_{6}-A_{\beta}) 
\nonumber \\
& + & (6K_{16}) \int d^{2}x\sqrt{g}g^{\alpha\beta}
(\partial_{\alpha}\theta_{1}-A_{\alpha})(\partial_{\beta}\theta_{6}-A_{\beta}),
\label{TH}
\end{eqnarray}
in terms of the local bond-angle field $\theta_{6}({\bf u})$ and 
the tilt-angle field $\theta_{1}({\bf u})$.
The constants multiplied to the stiffnesses $(K_{1},K_{16},K_{6})$ are 
introduced to show the symmetry in the recursion relations which will
be shown later.
The gauge field $A_{\alpha}$ describes how the basis vector rotates
under parallel transport according to the Gaussian curvature of the surface
\cite{Park-96}.
For simplicity, we dropped the tilt-bond interaction potential.
Effects of the tilt-bond interaction near the fixed points may change the
phase diagram qualitatively and deserve further investigation.
\par
Thus we have the full Hamiltonian ${\cal H}={\cal H}_{\rm HC}+
{\cal H}_{\rm TH}$ to describe a fluctuating tilted-hexatic membrane.
We follow Park and Lubensky's treatment of the topological defects
on fluctuating surfaces \cite{jp-lub1}, and obtain the tilted-hexatic membrane
partition function
\begin{equation}
{\cal Z} = \int {\cal D}{\bf R}{\cal D}\phi_{1}{\cal D}\phi_{6}
e^{-{\cal L}}
\end{equation}
with the effective Hamiltonian which is the 2-field sine-Gordon
Hamiltonian coupled to each other via off-diagonal propagator and coupled to
the geometry fluctuations by linear coupling to the scalar curvature
${\cal R}$ which is the twice the Gaussian curvature ${\cal R}=2K$
\begin{eqnarray}
{\cal L} & = & \frac{1}{2} \int d^{2}x \nabla\phi_{\mu} {\bf M}^{-1}_{\mu\nu}
\nabla\phi_{\nu} + \frac{i}{2\pi} \int d^{2}x (\phi_{1}+\phi_{6}){\cal R}
\nonumber  \\
   &  & -\frac{2y_{1}}{a^{2}} \int d^{2}x \cos\phi_{1}
-\frac{2y_{6}}{a^{2}} \int d^{2}x \cos\phi_{6}
+{\cal H}_{\rm HC},
\label{eff-Ham}
\end{eqnarray}
where we set $\beta=(1/k_{B}T)=1$ and
\begin{equation}
{\bf M}_{\mu\nu} = \left( \begin{array}{cc}
                          K_{1} & K_{16} \\
                          K_{16} & K_{6}
                          \end{array}
                   \right),
\end{equation}
$a$ is the short-distance cutoff,
$y_{1}$ ($y_{6}$) is the fugacity of vortices (disclinations), and
$\phi_{1}$ ($\phi_{6}$) is the conjugate field to vortices (disclinations).
When $\sqrt{g}$ is expanded in terms of $h$, nonvanishing lowest order terms
in $h$ are irrelevant and $\sqrt{g}$ is dropped in Eq. (\ref{eff-Ham}).
In order to establish the RG recursion relations for 
the tilted-hexatic rigidities, $(K_{1},
K_{16}, K_{6})$, and the fugacities, $(y_{1}, y_{6})$, 
we study the renormalization of
the two-point vertex functions $\Gamma^{(2)}_{\phi_{\mu}\phi_{\nu}}(q)$
for the effective Hamiltonian in Eq. (\ref{eff-Ham}).
Using the sine-Gordon renormalization analysis by 
Park and Lubensky \cite{jp-lub2}, we find three
critical regions:
\par
\begin{figure}
\centerline{\psfig{figure=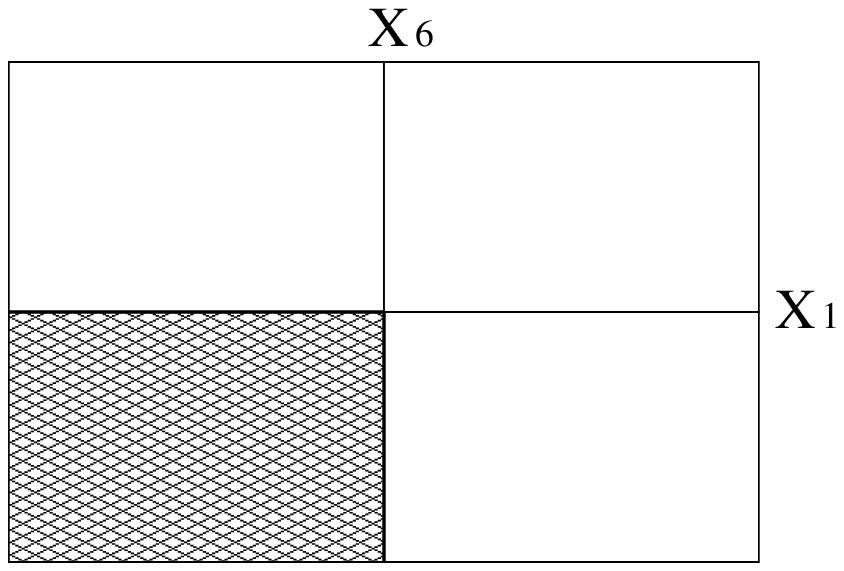}}
\caption{Phase diagram in the strong coupling regime near ${\cal S}_{\pm}
\equiv (K_{1},K_{16},K_{6})=(2/\pi,\pm 2/\pi,2/\pi)$.
The shaded region is the anti-locked (locked) tilted-hexatic phase, 
outside the disordered liquid phase. 
At the anti-locked (locked) tilted-hexatic/liquid phase boundary, 
disclinations and vortices unbind simultaneously.
}
\label{strong}
\end{figure}

1) the strong coupling regimes near the points ${\cal S}_{\pm} \equiv
(K_{1}, K_{16}, K_{6}) = (2/\pi, \pm 2/\pi, 2/\pi)$;
To leading order in the fugacities, we obtain 
\begin{eqnarray}
\frac{d}{dl} K_{1} & = & - 4\pi^{3} (K_{1}y_{1}+K_{16}y_{6})^{2}, \nonumber  \\
\frac{d}{dl} K_{6} & = & - 4\pi^{3} (K_{16}y_{1}+K_{6}y_{6})^{2}, \\
\frac{d}{dl} K_{16} & = & - 4\pi^{3} (K_{1}y_{1}+K_{16}y_{6})
(K_{16}y_{1}+K_{6}y_{6}),  \nonumber 
\end{eqnarray}
\begin{equation}
\frac{d}{dl} y_{1} = (2-\pi K_{1})y_{1}, \;\;
\frac{d}{dl} y_{6} = (2-\pi K_{6})y_{6}.
\end{equation}
In the above equations, $l$ is the renormalization group parameter.
As a check on these results, we set $K_{1} = K_{16}^{2}/K_{6}$ initially,
and find that this self-duality condition is preserved under our 
renormalization transformation;
\begin{equation}
\frac{d}{dl} \left( K_{1} - \frac{K_{16}^{2}}{K_{6}} \right) = 0
\end{equation}
as it should be.
To study the system in the critical regions near the points ${\cal S}_{\pm}$, 
it is useful to introduce deviations defined by
\begin{equation}
K_{1}^{-1}=\pi/2 (1 + X_{1}), \;\; K_{6}^{-1}=\pi/2 (1 + X_{6}),
\end{equation}
\begin{equation}
K_{16}^{-1}=\pi/2 (1 \pm X_{16}),
\end{equation}
as well as rescaled fugacities
\begin{equation}
Y_{1}^{2}=8\pi^{2}y_{1}^{2}, \;\; Y_{6}^{2}=8\pi^{2}y_{6}^{2}.
\end{equation}
To lowest order in these variables, the recursion relations become
\begin{equation}
\frac{dX_{1}}{dl} = \frac{dX_{16}}{dl} = \frac{dX_{6}}{dl} = 
(Y_{1}+Y_{6})^{2},
\end{equation}
\begin{equation}
\frac{dY_{1}}{dl} = 2X_{1}Y_{1}, \;\;
\frac{dY_{6}}{dl} = 2X_{6}Y_{6}.
\end{equation}
The flows generated by this system of the recursion relations are similar 
to the flows in the XY model \cite{XY-model}.
The phase diagram in these regimes is shown in Fig.~\ref{strong}.
In the shaded region, the long-wavelength properties of the
phase is described by the Hamiltonian in Eq.~(\ref{TH}) with
renormalized tilted-hexatic stiffnesses 
$(K_{1}^{\rm R},K_{16}^{\rm R},K_{6}^{\rm R})$. 
With the initial values $K_{1}(0)=K_{6}(0)=\pm K_{16}(0)$
and $Y_{1}(0)=Y_{6}(0)$, this effective Hamiltonian is minimized
when two angle fields are anti-locked (locked) by the constraint
$\nabla\theta_{1}=\mp 6\nabla\theta_{6}$ in the mean field level, and
taking into account thermal fluctuations
the resulting state is the anti-locked (locked) tilted-hexatic phase
near ${\cal S}_{+}$ (${\cal S}_{-}$)
in which the correlation functions show quasi-long-range order;
\begin{eqnarray}
D_{16} \equiv \langle \psi_{1}({\bf r})\psi_{6}(0) \rangle 
  & \simeq & r^{-\eta_{D}} \;\; {\rm near} \;\; {\cal S}_{+} \\
C_{16} \equiv \langle \psi_{1}({\bf r})\psi_{6}^{*}(0) \rangle 
  & \simeq & r^{-\eta_{C}} \;\; {\rm near} \;\; {\cal S}_{-},
\end{eqnarray}
where the exponents $\eta_{C}$ and $\eta_{D}$ are related by
\begin{equation}
\eta_{C}=\eta_{D}=\frac{1}{2\pi K_{1}^{R}}.
\end{equation}
Disclinations and vortices are irrelevant in this region. 
All $K_{1}$, $|K_{16}|$, and $K_{6}$ are destabilized and pushed toward
smaller values when either $y_{1}$ or $y_{6}$ starts to grow.
This happens when $K_{1} \leq 2/\pi$ or $K_{6} \leq 2/\pi$, and
we expect the anti-locked (locked) tilted-hexatic/liquid phase boundary when
disclinations and vortices unbind simultaneously.
\par
\begin{figure}
\centerline{\psfig{figure=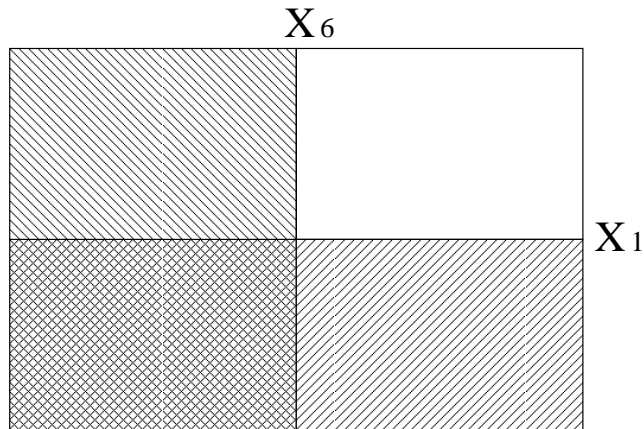}}
\caption{Phase diagram in the weak coupling regime near ${\cal W}
\equiv (K_{1},K_{16},K_{6})=(2/\pi,0,2/\pi)$.
Four phases-the unlocked tilted-hexatic phase (doubly shaded region), 
the tilted phase (shaded region with $X_{1}<0,X_{6}>0$),
the hexatic phase (shaded region with $X_{1}>0,X_{6}<0$), 
and the liquid phase (unshaded region)- are shown.
At the each phase boundary, 
disclinations and vortices unbind independently.
}
\label{weak}
\end{figure}

2) the weak coupling regime near the point ${\cal W} \equiv
(K_{1}, K_{16}, K_{6}) = (2/\pi, 0, 2/\pi)$;
To leading order in the fugacities, we obtain 
\begin{eqnarray}
\frac{d}{dl} K_{1} & = & - 4\pi^{3}(K_{1}^{2}y_{1}^{2}+K_{16}^{2}y_{6}^{2}), 
\nonumber  \\
\frac{d}{dl} K_{6} & = & - 4\pi^{3}(K_{16}^{2}y_{1}^{2}+K_{6}^{2}y_{6}^{2}), \\
\frac{d}{dl} K_{16} & = & - 4\pi^{3}(K_{1}K_{16}y_{1}^{2}+
K_{6}K_{16}y_{6}^{2})  \nonumber, 
\end{eqnarray}
\begin{equation}
\frac{d}{dl} y_{1} = (2-\pi K_{1})y_{1}, \;\;
\frac{d}{dl} y_{6} = (2-\pi K_{6})y_{6}.
\end{equation}
With the variables defined near the strong coupling fixed point except
for $K_{16}$ and defining $\bar{K}_{16}=(\pi/2)K_{16}$,
we rewrite the system near the point ${\cal W}$ to lowest order,
\begin{equation}
\frac{dX_{1}}{dl} = Y_{1}^{2} + \bar{K}_{16}^{2}Y_{6}^{2}, \;\;
\frac{dX_{6}}{dl} = Y_{6}^{2} + \bar{K}_{16}^{2}Y_{1}^{2}, 
\end{equation}
\begin{equation}
\frac{d\bar{K}_{16}}{dl} = - \bar{K}_{16} (Y_{1}^{2} + Y_{6}^{2}),
\label{K16}
\end{equation}
\begin{equation}
\frac{dY_{1}}{dl} = 2X_{1}Y_{1}, \;\;
\frac{dY_{6}}{dl} = 2X_{6}Y_{6}.
\end{equation}
Although the flows generated by this system are complicated, it is
easy to check that the quantity
\begin{equation}
{\cal C} = 2X_{1}^{2} + 2X_{6}^{2} - 4X_{1}X_{6}\bar{K}_{16}^{2} -
( Y_{1}^{2} + Y_{6}^{2} )
\label{C}
\end{equation}
is invariant to leading order along the trajectories,
\begin{equation}
\frac{d}{dl}{\cal C} = 0.
\end{equation}
Since ${\cal C}$ is entirely determined by Eq.~(\ref{C}) evaluated at
$l=0$, it is an analytic function of the initial conditions.
According to the recursion formula (\ref{K16}), the space $K_{16}=0$ is
attractive and the long wavelength properties are described by two
independent sets of the XY-like renormalization recursion relations;
one for the tilted-angle field, the other for the hexatic-angle field.
The phase diagram in this regime with $K_{16}=0$ is shown in Fig.~\ref{weak}.
$K_{1}$ is destabilized when $y_{1}$ starts to grow and this happens
when $K_{1} < 2/\pi$ and $K_{6}$ is destabilized when $y_{6}$ starts 
to grow and this happens when $K_{6} < 2/\pi$, respectively.
Since these instabilities occur independently, we expect there are
4 different phases; the unlocked tilted-hexatic phase, the hexatic phase,
the tilted phase, and the liquid phase.
The phase boundaries are given by $K_{1}=2/\pi$ and $K_{6}=2/\pi$.
\par
These two sets of  the recursion relations are different from those
derived by Nelson and Halperin.
Their recursion relations seem to be in range of the small gradient
cross coupling. Those are the same as the recursion relations
in the weak coupling regime in this Letter except the recursion relation
for the gradient cross coupling parameter.
They claimed $d K_{16}/dl =0$, and $K_{16}$ remains fixed at its initial
value to leading order in the fugacities.
However, we find the recursion relation for the gradient cross coupling
parameter nonvanishing in the weak coupling regime using the sine-Gordon 
field theoretic approach as well as the method of 
Kosterlitz \cite{Kosterlitz-74} employed by
Nelson and Halperin, and the space $K_{16}=0$ in the phase space is
attractive so that the initial nonvanishing value near the weak coupling
critical point is pushed toward smaller value.
In addition, we find the completely different set of the recursion relations
in the strong coupling regimes of the gradient cross coupling.
In Kosterlitz's way of description, configurations of hybrid 
vortex-disclination pairs are created in addition to vortex-antivortex and
$\pm 1$ disclination pairs to make the recursion relations in the strong
coupling regimes different from those in the weak coupling regime where
there is no logarithmic effective interaction between vortices and
disclinations.
Between the weak and strong coupling regimes along $K_{16}$ axis,
there must be the crossover regions near $K_{16}=\pm 1/\pi$ 
which are not accessible
by the perturbation expansion employed in this Letter.
Higher order terms in the fugacities and the deviations from the critical
points than the leading orders included in this Letter are necessary
in the crossover regions \cite{Amit-80}.
To connect the renormalization flows in a topologically correct way
in the weak and strong coupling regions, it is necessary to exploit the 
crossover region using the nonperturbative method.
\par
To complete the RG recursion relations for the effective Hamiltonian 
${\cal L}$, we study the renormalization of $\Gamma^{(2)}_{hh}(q)$ and obtain
\begin{equation}
\frac{d}{dl} \kappa = -\frac{3}{4\pi} \left[ 1-\frac{1}{4\kappa}
(K_{1} +2 |K_{16}| +K_{6}) \right]
\end{equation}
in the strong coupling regime and
\begin{equation}
\frac{d}{dl} \kappa = -\frac{3}{4\pi} \left[ 1-\frac{1}{4\kappa}
(K_{1} +K_{6}) \right]
\end{equation}
in the weak coupling regime.
The renormalization of the bending rigidity is the same in both regimes
except the appearance of $K_{16}$ in the recursion relation for the
strong coupling regime.
Defining $K_{1}+2K_{16}+K_{6}=k$ for the strong 
coupling regimes,
we find a fixed line corresponding to the crinkled
phase at $4\kappa = k$, and the crinkled-to-crumpled transition occurs
at $k=8/\pi$. The renormalized bending rigidity of the 
(anti-)locked tilted-hexatic
crinkled phase is $\kappa=2/\pi$.
In the weak coupling regime, there are three crinkled phases
differing from each other in the internal ordering.
The tilted and the hexatic crinkled phases has the renormalized bending 
rigidity $\kappa=1/2\pi$ and the unlocked tilted-hexatic crinkled phase
has $\kappa=1/\pi$.
Thus the crinkled phase in the strong coupling regime has the stiffer
bending rigidity than those in the weak coupling regime.
\par
We have presented phase transitions between the internally ordered phase to
disordered liquid phase on the fluctuating membranes using the
sine-Gordon renormalization analysis in the absence of the tilt-bond
interaction.
In this Letter we have considered two different kinds of internal orders;
tilt order and hexatic order. There may be other kinds of internal orders,
too. The results developed here can be applied to any two kinds of internal
orders. When there are two kinds of internal orders on the fluctuating
membrane, there exist two order-disorder transitions and the crumpling
transition. These order-disorder transitions occur simultaneously if the
coupling constant between two internal orders is in the strong coupling regime,
independently if it is in the weak coupling regime.
The crumpling transition occurrs when the 
last existing internal order disappears
and the membrane goes into the liquid phase.
Thus if the membrane has at least one kind of internal order,
the membrane is crinkled, not crumpled, 
with nonvanishing renormalized bending rigidity.
\par
Inclusion of the tilt-bond interaction in the tilted-hexactic Hamiltonian 
may change the quilitative properties of the fluctuating membranes
such as the stiffness of the membranes in the absence of the internal orders
or the relations between the local bond orientation and the local tilt
direction in the tilted-hexatic phase.
The tilt-bond interaction can be intepreted as the third internal order
with proper modification to make the renormalization
analysis complicated but possible. The sine-Gordon approach to the Hamiltonian
with the tilt-bond interaction is under investigation.
\par
The author is grateful to Prof. Lubensky for helpful conversations
and a careful reading of the manuscript.
This work was supported in
part by the Penn Laboratory for Research in the Structure of Matter under
NSF grant No.  91-20668.  

%\input{psfig}
%\newpage

\end{document}